\documentclass[artical]{IEEEtran}

\usepackage{amsmath}

\ifCLASSINFOpdf
\usepackage[pdftex]{graphicx}
  
\else
 
\fi

\begin{document}
%
\title{A Feedback Spin-Valve Memristive System}

\author{Weiran Cai,~\IEEEmembership{Student Member, IEEE}, Torsten Schmidt,
         Udo J\"orges, and  Frank Ellinger,~\IEEEmembership{Senior Member, IEEE}
\thanks{Weiran Cai, Udo J\"orges and Frank Ellinger are with the Chair for Circuit Design and Network Theory, and Torsten Schmidt is with the Chair for Fundamentals of Electronics, Faculty of Electronic and Information Engineering, Technische Universit\"at Dresden, Helmholtzstrasse 18, Barkhausenbau, 01069 Dresden, Germany. e-mail: weiran.cai@tu-dresden.de.}
}

\maketitle

\begin{abstract}
We propose theoretically a generalized memristive system based on controlled spin polarizations in the giant magnetoresistive material using a feedback loop with the classical Hall Effect. The dynamics can exhibit a memristive pinched hysteretic loop that possesses the self-crossing knot not located at the origin. Additionally, one can also observe a single-looped orbit in the device. We also provide a sufficient condition for the stability based on an estimation of the Floquet exponent. The analysis shows that the non-origin-crossing dynamics is generally permitted in a class of passive memory systems that are not subject to Ohm's Law. We further develop the prevailing homogeneous definition to a broadened concept of generalized heterogeneous memristive systems, permitting no self-crossing knot at the origin, and ultimately to the compound memory electronic systems.
\end{abstract}

\begin{IEEEkeywords}
memristor, nonlinear dynamics, spintronics, Hall Effect.
\end{IEEEkeywords}

\IEEEpeerreviewmaketitle

\section{Introduction}
\IEEEPARstart{I}{n} 1971, the concept of the memristor was coined by Leon O. Chua at Berkeley \cite{Chua}. But it was not until 2008 that S. Williams and his group at Hewlett-Packard Laboratories shed the first light on this scientific preemie with their TiO$_{\textrm{2-x}}$ memristor \cite{Stru}. It was also from then on that nonlinear electronics has been recognized as one of the comprising bases of the entire mansion of electronics. No later than this revival had Di Ventra, \textit{et al}, pushed the concept forward to a large family of passive memory systems, consisting of memristive, memcapacitive and meminductive systems \cite{Vent}. Deeper insights have also permeated into the territory of biophysiology, building links between neuronal transmissions, learning rules and memristive behaviours \cite{Lina}-\cite{Pers2}, making it an indeed interdisciplinary research subject. Parallel to the conceptual development, researchers have been enthused in seeking memristive systems on new physical mechanisms, at both the material level and the device level \cite{Schi}-\cite{Brat}. Among them is the spintronics of great interest, which has linked the memristive concept to amounts of established research results in this area. The spin Hall Effect in a semiconducting system with an inhomogeneous electron density has demonstrated a hysteretic pinch under periodically driving voltages \cite{Pers3}, and the spin diffusion and relaxation processes in the semiconductor/half-metal junction can also exhibit a memristive behaviour \cite{Pers4}. On the other hand, in the metal material, memristive systems are realizable by employing spin-torque-induced magnetization switching or magnetic-domain-wall motion \cite{Wang}\cite{Brat}. While people have been focused on other spin polarization mechanisms, in this paper we propose theoretically a memristive system based on the controlled spin polarization in the giant magnetoresistive material using a feedback loop with the classical Hall Effect. The system is of special interest in exhibiting hysteretic-pinched (as in a standard memristor) loop, which, however, has the self-crossing knot not at the origin. This characteristic has grown out of the prevailing categorization of memristive systems proposed by Di Ventra, \textit{et al}, but is still within a more general frame of memristive systems, as also suggested by Chua recently. It is hence defined as a generalized memristor model, which we name as the heterogeneous memristive system. The paper is organized as follows: we will introduce the generalized memristive systeml in Sec. II. The memristive dynamics is depicted and analyzed in Sec. III, including a sufficient condition for stability. In Sec. IV, we will relocate at the system's quasi-standard memristive dynamics, and discuss the generalization to the concept of heterogeneous memristive systems and compound memory systems. We will draw the conclusions in Sec. V.

\section{The Proposed Spin-Valve Memristive System}
\begin{figure}[!t]
\centering
\includegraphics[width=7.6cm]{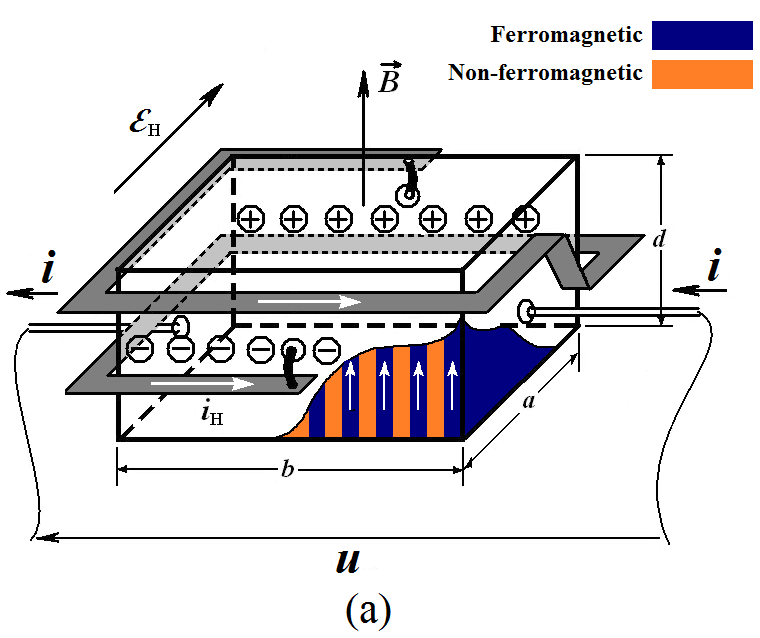}\\
\includegraphics[width=7.0cm]{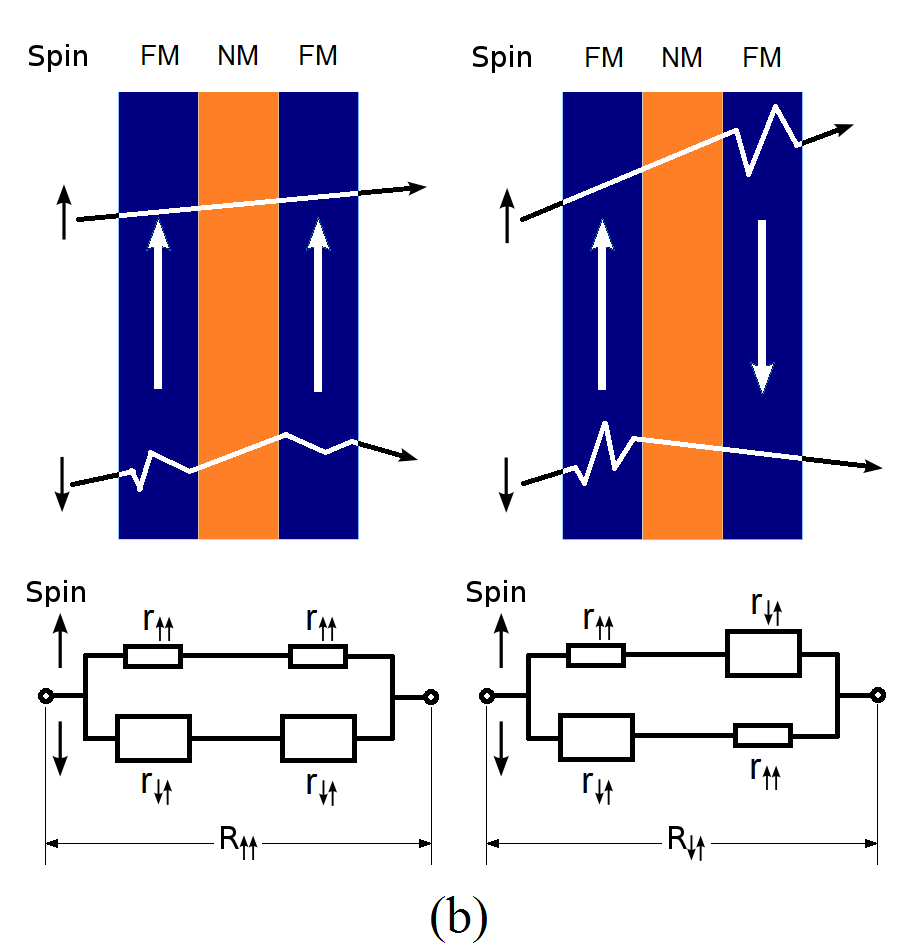}\\
\includegraphics[width=8.0cm]{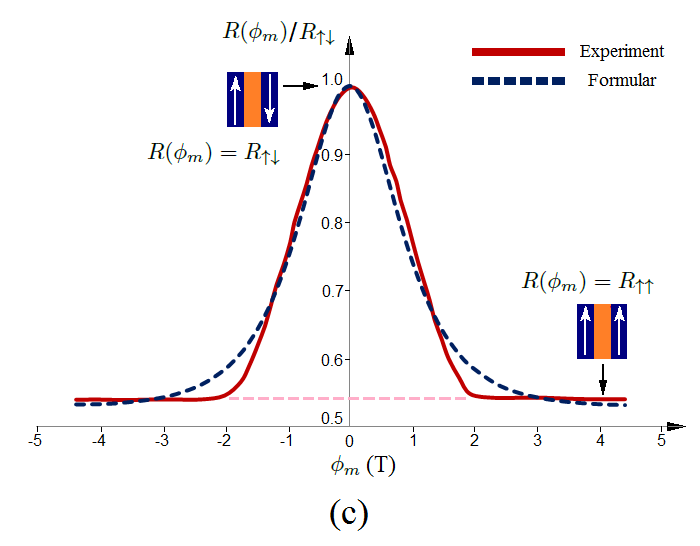}
\caption{The proposed feedback spin-valve memristor model in a giant magnetoresistive material of the current-perpendicular-to-plane (CPP) configuration. (a) structure: the spin in the ferromagnetic layers (red) are polarized adaptively to the temporally changing magnetic flux induced by the Hall current. (b) the principle of spin-valve in the GMR material and the equivalent circuit model. (c) the relation between the giant magnetoresistance $R(\phi_m)$ and the magnetic flux $\phi_m$ described by an empirical formula with a hyperbolic function, which is here used as a fitting curve to the experimental result in \cite{Baib}. The solid line is the experimental curve, and the dashed line is the fitting hyperbolic curve.  
}\label{fig1} 
\end{figure}

We propose a spin-valve memristive system established on a feedback loop. The loop controls the spin polarization in a giant magnetoresistive material via the classical Hall Effect, as shown in Fig. \ref{fig1}, which gives rise to a memory behaviour. Let us first consider a giant magnetoresistive (GMR) material in the form of stacked thin-films based on the spin-valve mechanism, of which the resistance reveals a strong dependence on the applied magnetic field \cite{Baib}\cite{Bina}. The principle of the significant change in the resistance is caused by the parallel or antiparallel spin polarizations: an electron passing through the GMR material will be scattered more if the spin of the electron is opposite to the direction of the magnetisation in the ferromagnetic layer, and the material hence expresses a larger resistance $r_{\uparrow\downarrow}$; otherwise, it expresses a smaller resistance $r_{\uparrow\uparrow}$. With this mechanism, an equivalent circuit for a stack of films can be constructed, so that the compound resistance under a magnetic field $R_{\uparrow\uparrow}$ (corresponding to parallel spin alignment in the material) is much lower than that without the magnetic field $R_{\uparrow\downarrow}$ (corresponding to unparallel spin alignment in the material). For our interest in the study of the dynamics, we can characterize the resistance empirically by a hyperbolic function of the magnetic flux $\phi_{m}$ as
\begin{equation}\label{r_phim}
R(\phi_m)=R_{\uparrow\uparrow}+(R_{\uparrow\downarrow}-R_{\uparrow\uparrow})~\textrm{sech} \left(\frac{\phi_m}{\phi_{m0}}\right)
\end{equation}
with $\phi_{m0}$ denoting a normalization quantity with the dimension of magnetic flux. The hyperbolic function is a fitting curve to the characteristic curves of a typical class of GMR materials. For example, as shown in Fig. \ref{fig1}, the hyperbolic curve (dashed line) given in Eq. \ref{r_phim} can well fit the experimental results (solid line) in the report of A. Fert, $\textit{et.al}$, where the thin-film structure is composed of alternating ferromagnetic and non-magnetic layers, namely, ...$|$Fe$|$Cr$|$Fe$|$Cr$|$..., with the thickness of Fe at 3 nm and Cr at 0.9$\sim$1.8 nm \cite{Baib}. On the other hand, the $sech$ function has a parabolic form when expanded to the second order for small fields, which has a decent match with the existing theorectical formalizations (for independent moment models) in \cite{Alli} and \cite{Elhi}. In spite of various GMR structures with specific quantum models, this empirical formalization deals with the problem of interest without losing generality. The characteristic curves of GMR materials are also not limited to the form given in \cite{Baib}, but they will generally give similar memristive dynamics in our proposed structure. 

To give rise to a memristive effect, we employ the classical Hall Effect to feed back a magnetic field as a control of the spin polarization in the ferromagnetic layers of the GMR material(the red-coloured layers in Fig. \ref{fig1}). The classical Hall Effect is based on a balance between the Lorenz force and the established inner electric field \cite{Hall}. The Hall electromotive force $\mathcal {E}_H$ is related to a magnetic field on the conductor by
\begin{equation}\label{eh_1}
\mathcal {E}_H=\frac 1 {ned} B i=\frac 1 {nedS} \phi_m i
\end{equation}
where $n$  is the carrier density in the magnetoresistive conductor, $e$  is the absolute value of the electron charge, $d$  is the transverse scale of the conductor, $S=ab$ is the horizontal cross-section area of the conductor. However, in this model, the magnetic field $\textbf{\textit{B}}$ is supplied internally by the Hall current $ i_H$. When the Hall current is drawn by the Hall electromotive force out to a coil, feeding back a controlling magnetic flux $\phi_m$ on the conductor itself, the balance is sustained in a dynamic way. The electromotive force equals the voltage drop on the coil 
\begin{equation}\label{eh_2}
\mathcal {E}_H=\frac{d\phi_m}{dt} + R_L i_H
\end{equation}
where $R_L$ is the coil resistance. It is to note that attributed to the ferromagnetic core, the inductance $L$ increases to a large extent with the enhanced permeability (by a factor of several hundred to thousand) and the dimension of the coil is allowed to be larger than that of the core due to the magnetic flux confinement. On the other hand, with the energy conservation, the entire system absorbs the power offered by an external voltage $u(t)$, equaling to the dissipation in the resistance of the conductor and the power in the coil:
\begin{equation}\label{ui}
ui=i^2 R(\phi_m) + \mathcal {E}_H i_H
\end{equation}
in which it is especially notable that due to the extraction of the Hall current, Ohm's law does not simply apply. In order to reveal the dominant factors contributing to the memristive dynamics, we base our modelling on the following approximations: 1.the magnetic flux changes with the Hall current $ i_H$ linearly through an inductance $L$ of the coil, \textit{i. e.}, $\phi_m=L\cdot i_H$. However, in terms of accuracy, it is worth pointing out that the ferromagnetic core saturation can cause a deviation from the linear relation at a high coil current. 2. the coil resistance $R_L$ is constant. In a more accurate manner, this resistance also includes a transverse resistance of the ferromagnetic conductor. For simplicity, we suppose that the transverse resistance is relatively small (by properly choosing the size ratio of the GMR material) and its temporal change is negligible in series with the main coil resistance, and hence $R_L$ is regarded constant. 3. other energy losses in the GMR material are not counted in for simplicity, which are mainly caused by the Eddy currents and the coercive losses (see, \textit{e. g.}, the GMR material in Ref. \cite{Baib} with a low coercitivity). These approximations aid to simplify the model, while also maintains the factors determining the dynamics, esp. the orbital topologies of our main interest. The nonlinearity in the inductance of the coil will limit the magnetic flux in practice. When these non-ideal effects are involved, they will deform the orbital shapes and modify the parameters into a more realistic range, but the system will not lose the memristive dynamics in general. Based on the above factors and approximations, the complete dynamics of the proposed memristive system can eventually be defined by the following equation set:
\begin{equation}\label{dm_a}
u(t) =  R(\phi_m) i +\frac{D_0} L \phi_m^2 
\end{equation}
with
\begin{equation}
R(\phi_m) = R_{\uparrow\uparrow}+(R_{\uparrow\downarrow}-R_{\uparrow\uparrow}) ~\textrm{sech}\left(\frac{\phi_m}{\phi_{m0}}\right)\nonumber
\end{equation}
and
\begin{equation}\label{dm_b}
\frac{d\phi_m}{dt} = D_0 \phi_m i - \frac{R_L}{L} \phi_m  
\end{equation}
in which $D_0$ denotes the coefficient $1/nedS$. It is supposed here that the average exertion time of spin torques under the magnetic field is sufficiently short compared to the time scale of the dynamics, and hence no time delay is involved in the above model. Special notice is to take on the extra term deviating to the resistive relation in Eq. (\ref{dm_a}), which is responsible for giving rise to a new memory effect beyond standard memristive systems. For material implementation, the current-perpendicular-to-plane (CPP) configuration of thin-films is to adopt (as shown in Fig. \ref{fig1}), for yielding the highest giant magnetoresistance along the external current \cite{Prat}-\cite{Kuma}, which possesses a coefficient $GMR \equiv (R_{\uparrow\downarrow}-R_{\uparrow\uparrow})/R_{\uparrow\uparrow}$  exceeding 1 at the present time.

\section{The Dynamics of the Memristive System}
\begin{figure}[!t]
\centering
\includegraphics[width=8.5cm]{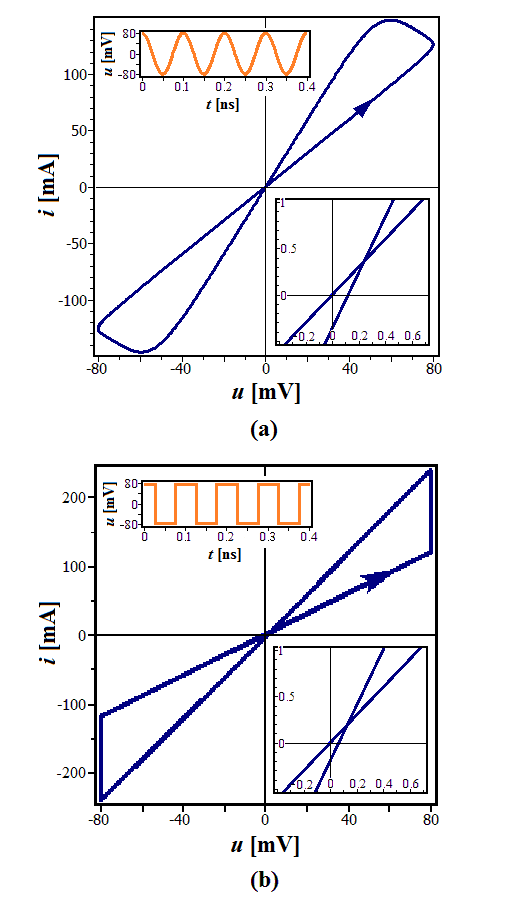}\\
\caption{The memristive dynamics of the spin-valve system, with (a) the sinusoidal input signal and (b) the square-wave input signal (hard switching response). The parameters are chosen as $e=1.6\times 10^{-19} ~\textrm{C}, n=10^ {20}/\textrm{cm}^3, R_{\uparrow\uparrow}=0.33 ~\Omega, R_{\uparrow\downarrow}=0.66~\Omega,  R_L=0.5~\Omega, a=1~\mu\textrm{m}, b=6~\mu\textrm{m}$, $d=1~\mu\textrm{m}$, $L=50~\textrm{nH}$ and $\phi_{m0}=30~\textrm{pWb}$. The sinusoidal and square-wave amplitudes are both $u_0=80~\textrm{mV}$ at the frequency $f=\omega/2\pi=10~\textrm{GHz}$. The zoomed-in display the deviation of the trajectory from the origin. The initial condition is set to be $\phi_m(0)=9~\textrm{pWb}$.  
}\label{fig2}
\end{figure}

\begin{figure}[!t]
\centering
\includegraphics[width=8.5cm]{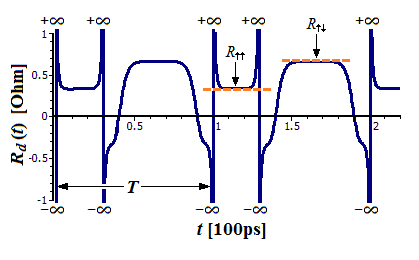}
\caption{The differential resistance $R_d(t)$ as a function of time, corresponding to the hysteretic loop in Fig. \ref{fig2}(a). The dashed lines denote the values of $R_{\uparrow\uparrow}$ and $R_{\uparrow\downarrow}$, which $R_d(t)$ approaches. $T$ denotes one period. 
}\label{fig3} 
\end{figure}

 The numerical calculations display that the proposed system can exhibit a memristive hysteretic pinch in the voltage-current phase plane when driven by a sinusoidal voltage signal $\left(u(t)=u_0\cos(\omega t)\right)$, as shown in Fig. (\ref{fig2}(a)), which realizes a memory effect of the dynamics produced by the feedback loop. However, unlike the typical hysteretic pinch in standard memristors, this generalized memristive system has its self-crossing knot of the orbit deviated from the origin of the phase plane, though the deviation can be rather small. This interesting characteristic is caused by the heterogeneous term in Eq. (\ref{dm_a}), in which the voltage $u$ is no loner a homogeneous equation about the current $i$ with the $L$-containing term. In this regarding, we will name such a system as a \textit{Heterogeneous Memristive System}, as a generalization of the standard memristive system, which we will name specifically as the \textit{Homogeneous Memristive System}. We will discuss in details the significance of this categorization in Sec. IV. For this specific system, it is worth pointing out that the proposed memristive system can be regarded as a standard memristive system in a limit. In fact, if we simplify Eq. (\ref{dm_a}) and (\ref{dm_b}) by diminishing the terms containing $L$ when these terms are small-valued (\textit{e.g.}, with a relatively large $L$) and integrating Eq. (\ref{dm_b}), we can recognize this degenerate model as a charge-controlled memristor with the memristance $M(q)=R(\phi_m)$, by taking $\phi_m^2$ as the state-variable: 
  \begin{subequations}\label{sm}
    \begin{alignat}{2}
      &u(t) =  R(\sqrt{\phi_m^2}) i \label{sm_a}\\
     &\frac{d\ln(\phi_m^2/\phi_{m0}^2)}{dt}=2D_0 i  \label{sm_b}
    \end{alignat}
  \end{subequations}
noting that $R(\phi_m)$ is an even function of $\phi_m$. Henceforth, the proposed memristive system can be regarded as charge-controlled memristor in this limit. Hard switching also demonstrates a hysteretic loop, as shown in Fig. \ref{fig2}(b).  

Regarding the local properties, we use the instantaneous differential resistance $R_d(t)\equiv du/di=(du/dt)\cdot (di/dt)^{-1}$ to characterize the orbits, as shown in Fig. \ref{fig3}. The behavior of $R_d(t)$ is a compound expression of both the dynamical process of the giant magnetoresistance and the non-Ohmic effect. We divide the $R_d(t)$ into two sections: the upward section $R_d^+(t)\equiv R_d(t\mid u(t):-u_0\rightarrow +u_0)$ and the downward section $R_d^-(t)\equiv R_d(t\mid u(t):+u_0\rightarrow -u_0)$. The differential resistance of the upward section at the orgin $R_d^+(t_o)$ approximates to $R_{\uparrow\downarrow}$ (when $u(t_o)= 0$ and $d\phi_m(t_o)/dt\approx 0$, $\phi_m(t_o)\approx 0$), while the downward section $R_d^-(t)$ approaches $R_{\uparrow\uparrow}$ at the self-crossing knot $t_c$. 

We have also observed that the system can also evolve to a memory system, which expresses a single-looped orbit (see Fig. (\ref{fig2}(b)). Such orbit rises up when the $L$-containing terms gain dominancy in Eq. (\ref{dm_a}) and (b). In contrast to the standard memristive systems, the single-looped orbit also has a part of the trajectory that does not cross the origin when the voltage is zero. This is also simply attributed to the additional non-zero term to the resistive relation in Eq. (\ref{dm_a}). The hard switching also demonstrates a single-looped orbit, as shown in Fig. \ref{fig2}(c).

\begin{figure}[!t]
\centering
\includegraphics[width=8cm]{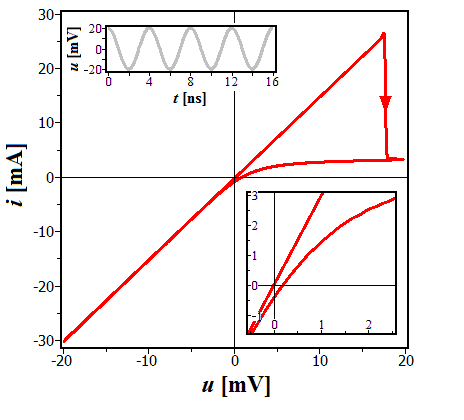}\\
\caption{The single-looped dynamics of the spin-valve system, with a sinusoidal input signal. The parameters are chosen as $e=1.6\times 10^{-19} ~\textrm{C}, n=10^ {20}/\textrm{cm}^3, R_{\uparrow\uparrow}=0.33 ~\Omega, R_{\uparrow\downarrow}=0.66~\Omega,  R_L=0.5~\Omega, a=1~\mu\textrm{m}, b=6~\mu\textrm{m}$, $d=1~\mu\textrm{m}$, $ L=4~\textrm{nH}$ and $\phi_{m0}=5~\textrm{pWb}$. The sinusoidal amplitudes is $u_0=20~\textrm{mV}$ at the frequency $f=\omega/2\pi=0.25~\textrm{GHz}$. The zoomed-in display the deviation of the trajectory from the origin. The initial condition is set to be $\phi_m(0)=9~\textrm{pWb}$.
}\label{sg_loop} 
\end{figure}

Globally, the bi-looped and single-looped orbits are topologically homeomorphic to the Lissajous figures for the ratio of 1 and 2, respectively. In fact, if we look at the dynamics in the $u-\phi_m$ phase plane, the two orbits are both homeomorphic to a circle. The uprising of the single-looped orbit is because the terms containing $L$ gain dominancy in Eq. (\ref{dm_a}) and (b).  

\begin{figure}[!t]
\centering
\includegraphics[width=7.8cm]{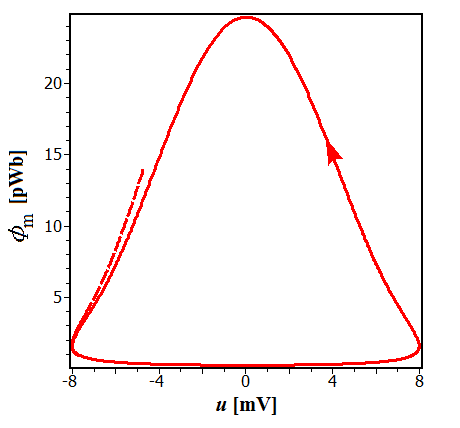}
\caption{The limit cycle in the $u-\phi_m$ plane, with the parameters given in Fig. \ref{fig2}. The dashed line represents a part of the transient.
}\label{fig4} 
\end{figure}

We are interested now in the condition for stability. Given proper parametric conditions, this proposed memristive systems can display a stable behavior after experiencing a transient. It is observable that the dynamics of the model in the $u-\phi_m$ phase plane displays a one-looped limit cycle under a sinusoidal driving signal, \textit{e. g.}, as displayed in Fig. \ref{fig4}. But to confirm it as an attractor, it is necessary to know the parameter range for stability. The model has one state-variable, as rewritten in the following form 
  \begin{subequations}\label{st}
    \begin{alignat}{2}
     \frac{d\phi_m}{dt} &= \frac{D_0}{R(\phi_m)} \phi_m u(t) - \frac{D_0^2}{R(\phi_m)L} \phi_m^3-\frac{R_L}{L} \phi_m \label{st_a}\\
     u(t) &= u_0\cos(\omega t) \label{st_b}
    \end{alignat}
  \end{subequations}
We here use the Poincar\'{e} map to study the stability of the periodic $u-\phi_m$ planar orbit \cite{Stro}. The method converts a problem about a closed orbit into a problem about the fixed points of a mapping. However, the rare possibility of achieving an analytical expression of the mapping constrains most of on-hand analysis on numerical calculations for specific parameters. In order to see the parameter range for stability, we here rather provide a sufficient condition for the asymptotic stability by estimating the Floquet exponent of the system.

\begin{figure}[!t]
\centering
\includegraphics[width=7.8cm]{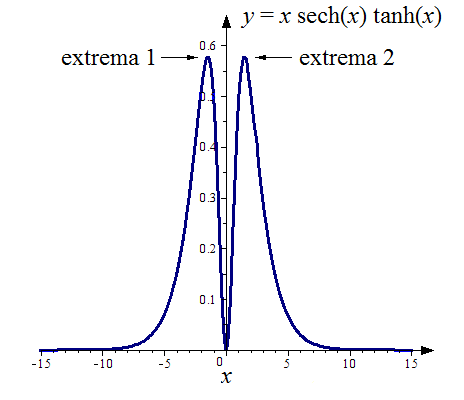}
\caption{The function $y(x)=x\textrm{sech}(x)\textrm{tanh}(x)$ plotted numerically. It has two extremas, \textit{i.e.}, $(x_{max},y_{max})=(\pm 1.463, 0.577)$, which are symmetric about the $y$ axis. The parameter $\Gamma_0$ is defined as the maximum value of the function.
}\label{gam} 
\end{figure}

Let $\phi_m^*(t)$  be the closed orbit and $\eta(t)$ be an infinitesimal perturbation: $\phi_m(t)=\phi_m^*(t)+\eta(t)$. Substituting this relation into Eq. (\ref{st_a}) and using the Taylor expansion to the first order of $\eta$, we can rewrite the equation in the following form:
\small\begin{equation} \label{sub}
\begin{split}
     &\frac{d(\phi_m^*+\eta)}{dt} = \Big\{\Big[ D_0(R_{\uparrow\downarrow}-R_{\uparrow\uparrow})\frac{\phi_m^*}{\phi_{m0}} \textrm{sech} \left(\frac{\phi_m^*}{\phi_{m0}}\right) \tanh \left(\frac{\phi_m^*}{\phi_{m0}}\right) \\
     &~+D_0R(\phi_m^*)\Big]\frac{u(t)}{R(\phi_m^*)^2}
-\Big[ D_0^2(R_{\uparrow\downarrow}-R_{\uparrow\uparrow})\frac{\phi_m^{*3}}{\phi_{m0}} \textrm{sech} \left(\frac{\phi_m^*}{\phi_{m0}}\right)\\
     &~~~\cdot \tanh \left(\frac{\phi_m^*}{\phi_{m0}}\right)+3D_0^2\phi_m^{*2}R(\phi_m^*)\Big]\frac{1}{R(\phi_m^*)^2L} -\frac{R_L}{L}\Big\}\eta\\
     &+\frac{D_0}{R(\phi_m^*)} \phi_m^* u(t) - \frac{D_0^2}{R(\phi_m^*)L} \phi_m^{*3}-\frac{R_L}{L} \phi_m^* 
\end{split}
\end{equation}\normalsize
Since $\phi_m^*(t)$ is a solution of Eq. (\ref{st_a}), the following relation should hold
\begin{equation}\label{sol_sp}
\frac{d\phi_m^*}{dt} = \frac{D_0}{R(\phi_m^*)} \phi_m^* u(t) - \frac{D_0^2}{R(\phi_m^*)L} \phi_m^{*3}-\frac{R_L}{L} \phi_m^*
\end{equation}
Substracting Eq. (\ref{sub}) by Eq. (\ref{sol_sp}), we can arrive at the equation about the error function $\eta(t)$, given by
\begin{equation}\label{eta}
\frac{d\eta(t)}{dt}=\mathcal{S}(t)\eta(t)
\end{equation}
where
\begin{equation} \label{s}
\begin{split}
     &\mathcal{S}(t) = \Big[ D_0(R_{\uparrow\downarrow}-R_{\uparrow\uparrow})\frac{\phi_m^*}{\phi_{m0}} \textrm{sech} \left(\frac{\phi_m^*}{\phi_{m0}}\right) \tanh \left(\frac{\phi_m^*}{\phi_{m0}}\right) \\
     &~+D_0R(\phi_m^*)\Big]\frac{u(t)}{R(\phi_m^*)^2}
-\Big[ D_0^2(R_{\uparrow\downarrow}-R_{\uparrow\uparrow})\frac{\phi_m^{*3}}{\phi_{m0}} \textrm{sech} \left(\frac{\phi_m^*}{\phi_{m0}}\right)\\
     &~~~\cdot \tanh \left(\frac{\phi_m^*}{\phi_{m0}}\right)+3D_0^2\phi_m^{*2}R(\phi_m^*)\Big]\frac{1}{R(\phi_m^*)^2L} -\frac{R_L}{L}
\end{split}
\end{equation}
In order to calculate the evolution of the perturbation on the surface of section after each period $T=2\pi/\omega$, we integrate over $t$  from $nT$ to $(n+1)T$. It comes to
\begin{equation}\label{eta_T}
|\eta\left((n+1)T\right)|=e^{\lambda} | \eta( nT)|
\end{equation}
where   
\begin{equation}\label{lam}
\lambda=\int_0^T \mathcal{S}(t) dt
\end{equation}                                                                                                                                  
is the Floquet exponent of the linearized Poincar\'{e} map. Let us now estimate the $\lambda$ without knowing the analytical solution $\phi_m^*(t)$. The even function $\phi_m^*/\phi_{m0}\cdot \textrm{sech}(\phi_m^*/\phi_{m0}) \tanh(\phi_m^*/\phi_{m0})$ in the integral is actually a positive bounded function in $\phi_m^*$, having 0 as the minimum value and $\Gamma_0$ as the maximum value, where $\Gamma_0\approx 0.577$ denotes a positive real value at its two symmetric extrema (see Fig. (\ref{gam})); also is $\phi_m^{*3}/\phi_{m0}\cdot \textrm{sech}(\phi_m^*/\phi_{m0}) \tanh(\phi_m^*/\phi_{m0})$ positive bounded. Therefore, the two parts in the integrand are bounded: 
\begin{subequations}\label{bound}
\small  \begin{equation}\label{b1}
  \begin{split}
    M_1  &\equiv  \frac{1}{R_{\uparrow\downarrow}}\leq \Big[(R_{\uparrow\downarrow}-   R_{\uparrow\uparrow})\frac{\phi_m^*}{\phi_ {m0}}\textrm{sech}\left(\frac{\phi_m}{\phi_{m0}}\right)\tanh\left(\frac{\phi_m}{\phi_{m0}}\right) \\                
+& R(\phi_m^*) \Big]\frac{1}{R(\phi_m^*)^2}\leq  \frac{(R_{\uparrow\downarrow}-R_{\uparrow\uparrow})\Gamma_0+R_{\uparrow\uparrow}}{R_{\uparrow\uparrow}^2} \equiv M_2\\
  \end{split}
  \end{equation}\normalsize
and
\begin{eqnarray}\label{b2}
    &&D_0^2 \Big[ (R_{\uparrow\downarrow}-R_{\uparrow\uparrow})\frac{\phi_m^{*}}{\phi_{m0}}\textrm{sech}\left(\frac{\phi_m^*}{\phi_{m0}}\right)\tanh\left(\frac{\phi_m^*}{\phi_{m0}}\right) \nonumber\\
&&~~~+3R(\phi_m^*) \Big] \frac{\phi_m^{*2}}{R(\phi_m^*)^2} \geq 0  
  \end{eqnarray}
\end{subequations}             

Noting that $u(t)=u_0\cos(\omega t)$ is alternatingly positive and negative in half-periods, \textit{i.e.}, $0\leq u(t)\leq u_0$, for $t\in [0,T/4]\cup [3T/4,T]$ and $-u_0\leq u(t)\leq 0$, for $t\in [T/4,3T/4]$, we integrate and estimate Eq. (\ref{lam}) in three intervals with the relations in Eq. (\ref{b1}) and (\ref{b2}). This can lead to the following inequation:
\begin{equation}\label{lamb_in}
 \begin{split}
\lambda &\leq D_0 M_2 \left( \int_0^{T/4}+\int_{3T/4}^T \right) u(t)dt + D_0 M_1 \int_{T/4}^{3T/4} u(t)dt  \\
& -\frac{R_LT}L 
=\frac{2 u_0 D_0 (M_2-M_1)}{\omega}-\frac{2\pi R_L}{\omega L}
 \end{split}
\end{equation}             
Therefore, when the following condition is satisfied,
\begin{eqnarray}\label{st_con}
D_0 \left( \frac{(R_{\uparrow\downarrow}-R_{\uparrow\uparrow})\Gamma_0+R_{\uparrow\uparrow}}{R_{\uparrow\uparrow}^2}-\frac{1}{R_{\uparrow\downarrow}} \right) < \frac{\pi R_L}{u_0L}
\end{eqnarray}     
the Floquet exponent is guaranteed negative, which indicates that the system is asymptotically stable (a stable limit cycle). This condition is frequency independent. The left hand side of Eq. (\ref{st_con}) is determined solely by the GMR material, while the right hand side is determined by the feedback coil and the input signal.

\section{Discussions}
1. The spin-valve model as a standard memristor

Let us relocate at the degenerate model (with the $L$-containing terms being non-dominant). Being both charge-controlled memristor models, we find it especially interesting that the spin-valve memristive system has a close analogy to the often referred TiO$_{\textrm{2-x}}$ memristor of S. Williams, \textit{et al} \cite{Stru}. To see this, we expand the \textit{sech} term in Eq. (\ref{r_phim}) to the second order (valid for small fields: $\textrm{sech}(x)=1-x^2/2+O(x^4)$) and compare the degenerated equation set with that of the TiO$_{\textrm{2-x}}$ model: 
  \begin{subequations}\label{com}
    \begin{alignat}{3}
     u(t) = R(\phi_m)  i ~~~ &\longleftrightarrow~~~ u(t) = R(w)  i \label{com_a}\\
\frac{d\ln(\phi_m^2/\phi_{m0}^2)}{dt} = \frac{2}{nedS} i ~~~&\longleftrightarrow~~~ \frac{dw}{dt} = \frac{\mu_V R_{ON}}{D} i  \label{com_b}\\
R(\phi_m) = R_{\uparrow\downarrow}-(&R_{\uparrow\downarrow}-R_{\uparrow\uparrow})\frac{\phi_m^2}{2\phi_{m0}^2}~ \sim \phi_m^2 \nonumber\\
 &\longleftrightarrow&~~\nonumber\\
 R(w)=R_{OFF}-&(R_{OFF}-R_{ON})\frac w D~\sim w  \label{com_c}
    \end{alignat}
  \end{subequations}
Apparently, for small $\phi_m$, taking $\phi_m^2$  as the state-variable, they have general similarities in the function relations, except a logarithmic function in the former. However, they have a complete distinction in their physical mechanisms. They differ also in the resistance range, being a low conductive resistance for the spin-valve model. Furthermore, it is notable that opposite to the TiO$_{\textrm{2-x}}$ memristor, the spin-valve memristor does not have a nonvolatile memory when switched off, as the current in the coil cannot hold permanently.  
\\

2. Generalized memristive systems

Since Di Ventra, \textit{et al}, generalized the concept of memristor of Chua, to memristive, memcapacitive and meminductive systems, any newly born elementary passive memory system has been seeking a belonging category in the form of 
  \begin{subequations}\label{msys1}
    \begin{alignat}{2}
     y(t)&=g( \mathbf{x},z,t)\cdot z(t) \label{msys1_a}\\
 \mathbf{\dot x}&= \mathbf{f}( \mathbf{x},z,t)  \label{msys1_b}
    \end{alignat}
  \end{subequations}
or of its variations, describing two complementary constitutive variables (current, charge, voltage or flux), where $g$ is a generalized response, and $\mathbf{f}$ is a vector function, with $y$ being the input signal, $z$ the output signal, and $\mathbf{x}$ the state-variable vector \cite{Vent}. It has been successful for a very large class of memristive device structures \cite{Stru}\cite{Schi}-\cite{Brat}. This formal description, however, is based on the linear fundamental electronic elements, \textit{i.e.}, the resistor, capacitor and inductor, which in common has a homogeneous characteristic that the trajectory must cross the origin in the $u-i$ phase plane, due to Eq. (\ref{msys1_a}). We are curious about the necessity of holding this strong constraint for an arbitrary non-combinatorial passive memory system beyond the above. Apparently, our proposed spin-valve systeml has grown out of the prevailing categorization of memristive systems, which resembles a standard memristive system (pinched with self-crossing at the origin). In fact, the non-origin-self-crossing characteristic is permitted in a class of systems that are not subject to the Ohm's law. On the other hand, any passive system that can express a memory effect should be regarded as a memristive system in a general sense. In \cite{Chua_AP}, L. Chua also concluded such property of resistance switching memory devices as: "\textit{If it's pinched, it's a memristor.}" This implies that the memory effect is the essential property of a memristor, while the origin-crossing property holds only for a class of memristors. In this context, it leads us to the generalized description of a heterogeneous memristive system:
  \begin{subequations}\label{msys2}
    \begin{alignat}{2}
     u(t)&=g( \mathbf{x},i,t)\cdot i(t) +\widetilde{g}( \mathbf{x},i,t) \label{msys2_a}\\
    \mathbf{\dot x}&= \mathbf{f}( \mathbf{x},i,t)  \label{msys2_b}
    \end{alignat}
  \end{subequations}
which does not require a solution for $\{i(t_0)=0, \widetilde{g}\left(  \mathbf{x}(t_0),0,t_0 \right)=0\}$. Such systems are of special interest to exhibit hysteretic loop with the self-crossing knot not located at the origin of the $u-i$ phase plane for the non-zero heterogeneous term $\widetilde{g}( \mathbf{x},i,t)$ in Eq. (\ref{msys2_a}). Heterogeneous memcapacitive and meminductive systems are also permitted for existence and can be defined in the same manner. 

This generalization can lead to a further broadened concept, as we have noticed that when the heterogeneous term $\widetilde{g}( \mathbf{x},i,t)$ gains sufficient dominancy, the dynamics of the system may express characteristics beyond a standard memristive system. This implies that intrinsic passive memory systems can possess more than one attributes. These attributes are, however, so physically entangled to each other that the boundaries of memristive (resistive), memcapacitive (capacitive) and meminductive (inductive) systems are vague, and hence they should have a compound form in general
  \begin{subequations}\label{msys3}
    \begin{alignat}{2}
    & \mathbf{H}(\mathcal{V})=  \mathbf{0} \label{msys3_a}\\
    & \mathbf{\dot x}= \mathbf{F}(\mathcal{V})  \label{msys3_b}
    \end{alignat}
  \end{subequations}
where $\mathbf{H}$ and $\mathbf{F}$ are two vectors of implicit algebraic functions of a set of variables $\mathcal{V}$:  
\begin{eqnarray}\label{v}
\mathcal{V}\equiv \left\{u,i,u^{(n)},i^{(m)},\int^{(k)}udt, \int^{(l)}idt, \mathbf{x},t\mid n,m,k,l \in \mathcal{N} \right\}.\nonumber
\end{eqnarray}
Due to the shifted boundaries, we would refer to such a system simply as a compound memory electronic system (or a \textit{memtronic} system in short). It is foreseeable that some complex nano-, organic or biological systems possessing electric memory have to be described in such a general manner.

\section{Conclusions}
We have proposed theoretically a generalized memristive system based on the feedback spin-valve mechanism in the giant magnetoresistive material with the classical Hall Effect. The dynamics can exhibit a pinched hysteretic loop that possesses the self-crossing knot not located at the origin of the phase plane, which is of its uniqueness with regard to standard memristive systems. We have also observed that a single-looped orbit is allowed for this system, when the heterogeneous terms in the governing equations gain sufficient dominancy. Our analysis shows that such non-origin-crossing dynamics is generally permitted in a class of passive memory systems that are not subject to Ohm's Law, which has grown out of the prevailing homogeneous categorization of memristive systems. We have hence introduced a broadened concept of the heterogeneous memristive systems and ultimately the compound memory electronic systems. For the stability of the proposed model, we have provided a sufficient condition based on an estimation of the Floquet exponent in the linearized Poincar\'{e} map.

\ifCLASSOPTIONcaptionsoff
  \newpage
\fi




\end{document}